\documentclass[twocolumn,showpacs,showkeys,preprintnumbers,amsmath,amssymb]{revtex4}
 \usepackage{dcolumn}
 \usepackage{bm}
 \usepackage{graphicx}
 \usepackage{subfigure}
 \begin{document}

 \title{Time domain analysis of superradiant instability for the charged stringy black hole-mirror system}

 \author{Ran Li$^1$}
 \email{021149@htu.cn}

 \author{Yu Tian$^{2,3}$}
 \email{ytian@ucas.ac.cn}

 \author{Hongbao Zhang$^4$}
 \email{hzhang@vub.ac.be}

 \author{Junkun Zhao$^1$}
 \email{zhaojkun1991@163.com}

 \affiliation{
 $^1$ Department of Physics, Henan Normal University, Xinxiang 453007, China\\
 $^2$ School of Physics, University of Chinese Academy of Sciences, Beijing 100049, China\\
 $^3$ State Key Laboratory of Theoretical Physics, Institute of Theoretical Physics,
 Chinese Academy of Sciences, Beijing 100190, China\\
 $^4$ Theoretische Natuurkunde, Vrije Universiteit Brussel, and The International Solvay Institutes,\\
 Pleinlaan 2, B-1050 Brussels, Belgium}

 \begin{abstract}

  It has been proved that the charged stringy black holes are stable under the perturbations of massive charged scalar fields. However, superradiant instability can be generated by adding the mirror-like boundary condition to the composed system of charged stringy black hole and scalar field. The unstable
  boxed quasinormal modes have been calculated by using both analytical and numerical method.
  In this paper, we further provide a time domain analysis by performing a long time evolution of charged scalar field configuration in the background of the charged stringy black hole with the mirror-like boundary condition imposed. We have used the ingoing Eddington-Finkelstein coordinates to derive the evolution equation, and adopted Pseudo-spectral method and the forth-order Runge-Kutta method to evolve the scalar field with the initial Gaussian wave packet. It is shown by our numerical scheme that Fourier transforming the evolution data coincides well with the unstable modes computed from frequency domain analysis. The existence of the rapid growth mode makes the charged stringy black hole
  a good test ground to study the nonlinear development of superradiant instability.

 \end{abstract}



 \maketitle

 \section{introduction}

 Superradiance is an interesting classical effect of the rotating or charged black holes
 \cite{zeldovich,bardeen,misner,starobinsky}. Although in any classical process in black hole spacetime
 no particle (even for photon) can escape from the event horizon of black hole, the energy of rotating or charged black hole can be extracted out via the superradiance process. It should also be noted that this classical superradiance process does not violate the area theorem. Consider, for example, the superradiance effect of rotating black hole. Suppose a bosonic wave of the form $e^{-i(\omega t-m\phi)}$, with $\omega$ and $m$ being the frequency and azimuthal quantum number, is scattered by the event
 horizon. If $Re[\omega]<m \Omega_H$ with $\Omega_H$ being the angular velocity of event horizon, the incident wave gets amplified by the black hole rotating. For the charged black holes, the charged bosonic fields should be considered in order to create this effect, where the charge of perturbation field play the role of rotating \cite{bekenstein}.

 This effect may trigger an instability if the wave repeatedly scatters off the black hole, which is named as superradiant instability or black hole bomb mechanism \cite{press,cardoso2004bomb}. This can be achieved by either introducing a mass term of the bosonic field or adding a mirror-like boundary condition outside the black hole, which will reflect the bosonic wave repeatedly. Then the energy extracted from the black hole can grow exponentially with the time. For many cases, this instability triggered by superradiance has been confirmed by both analytical and numerical methods \cite{Rosa,Lee,leejhep,jgrosa,hod2013prd,hodbhb,kerrunstable,detweiler,strafuss,dolan,Hod,hodPLB2012,
 konoplyaPLB,DiasPRD2006,zhangw,cardoso2004ads,cardoso2006prd,KKZ,aliev,uchikata,rlplb,zhang,
 knopolya,rlepjc,clement,randilaton}. One can refer to Ref.\cite{superradiance} for a recent review on this topic. However, this is not the full story. It is shown that, in some cases, the scalar fields do behave the superradiance effect, while the instability does not occur. For example, the Reissner-Nordstrom black hole is stable against the superradiance\cite{hodrnplb2012,hodrnplb2013,hodrnprd2015,gz}.

 In \cite{liprd}, we have shown that the mass term of charged scalar field can not support quasi-bound state outside the charged stringy black hole. This is to say that the charged stringy black hole is stable against the massive charged scalar perturbation. This result was further confirmed in \cite{binwang} by
 computing the quasinormal modes of charged scalar perturbation.
 In \cite{liepjc2014}, we have studied the superradiant instability of the scalar field in the background of charged stringy black hole due to a mirror-like boundary condition. The analytical expression of the unstable superradiant modes is derived by using the asymptotic matching method \cite{cardoso2004bomb}. Soon later, this analytical result is further confirmed by the recent numerical studies \cite{liplb2015}. Motivated by the work \cite{raduprl}, the stationary cloud-like configurations of scalar field satisfying the superradiant critical frequencies are also studied in \cite{liepjc2015}. However, the instability is only revealed by computing the unstable boxed quasinormal modes in the frequency domain analysis \footnote{It should be noted that
 the boxed quasinormal mode is also called as quasi-bound state in literature.}.

  In this paper, we shall perform the time domain analysis of the charged scalar field configuration in the above charged stringy black hole background. Although the time domain analysis of superradiant instability is a challenging work due to the small growth rate, a time domain study of superradiant instability for the scalar field on Kerr spacetime has been initiated in \cite{dolanprd2013}. Instead of the usually used Schwarzschild like coordinates, we will exploit the ingoing Eddington-Finkelstein coordinates, where the partial differential evolution equation is generically first order in time such that there is no need for one to introduce an auxiliary field for numerical computation. In addition, compared to the finite difference method, the pseudo-spectral method has become an increasingly popular numerical technique to study the black hole physics in particular due to its higher computation efficiency. For example, the pseudo-spectral method has recently been used to calculate the quasinormal modes of asymptotically flat black holes in \cite{cook} and to study the gravitational wave collapse in \cite{Hilditch}. One can refer to Ref.\cite{spectral} for a review of this numerical method. With this in mind, we shall resort to the pseudo-spectral method in the spacial direction supplemented with the forth order Runge-Kutta method in time to numerically evolve an initial Gaussian wave packet. As a result, the unstable modes, obtained from the Fourier transforming the evolution data, prove to be in good agreement with those computed from the frequency domain analysis. We also confirm the existence of the rapid growth modes, which makes the charged stringy black hole a good test ground to study the nonlinear development of superradiant instability.

 This paper is organized as follows. In Sec. II, we describe the basic setup of a charged scalar field with mirror-like boundary condition in charged stringy black hole background.  The involved partial differential equation that governs the evolution is also derived. In Sec. III, our numerical scheme is detailed. We present our numerical results and relevant discussions in Sec. IV. Final remarks are made in the last Section.

 \section{framework}

 The black hole background considered in the present paper is a static spherically symmetric charged black hole in the low energy effective theory of heterotic string theory in four dimensions, which was firstly found by Gibbons and Maeda in \cite{GM} as well as independently found by Garfinkle, Horowitz, and Strominger in \cite{GHS} a few years later. The metric is explicitly given by
 \begin{eqnarray}
 ds^2&=&-\left(1-\frac{2M}{r}\right)dt^2+\left(1-\frac{2M}{r}\right)^{-1}
 dr^2\nonumber\\
 &&+r\left(r-\frac{Q^2}{M}\right)(d\theta^2+\sin^2\theta d\phi^2)\;,
 \end{eqnarray}
 and the electric potential and dilaton field are given by
 \begin{eqnarray}
 &&A_t=-\frac{Q}{r}\;,\nonumber\\
 &&e^{2\Phi}=1-\frac{Q^2}{Mr}\;.
 \end{eqnarray}
 The parameters $M$ and $Q$ are the mass and electric charge of the black hole, respectively. The event horizon of black hole is located at $r=2M$ and the singularity is located at $r=Q^2/M$. We only consider the non-extremal case in which the parameters satisfy the condition $Q<\sqrt{2} M$.

 For the latter convenience of numerical calculation, we would like to adopt the ingoing null coordinate defined by
 \begin{eqnarray}
 v=t+r^*\;,
 \end{eqnarray}
 with $r^*$ being the tortoise coordinate
 \begin{eqnarray}
 r^*=\int \frac{dr}{1-\frac{2M}{r}}=r+2M\ln\left|\frac{r}{2M}-1\right|\;.
 \end{eqnarray}
 Then the metric in the ingoing Eddington-Finkelstein coordinates reads
 \begin{eqnarray}
 ds^2&=&-\left(1-\frac{2M}{r}\right)dv^2+2dvdr\nonumber\\
 &&+r\left(r-\frac{Q^2}{M}\right)(d\theta^2+\sin^2\theta d\phi^2)\;,
 \end{eqnarray}
 and the electric potential can be written as
 \begin{eqnarray}
 A_v=-\frac{Q}{r}\;,
 \end{eqnarray}
 where we have also performed a gauge transformation $A\rightarrow A-\nabla\psi$ with $\psi=Q\int \frac{dr}{r-2M}$.

 For simiplicity, it is sufficient to consider the test charged massless scalar field, the dynamics of which is governed by the Klein-Gordon equation
 \begin{eqnarray}
 (\nabla_\nu-iqA_\nu)(\nabla^\nu-iqA^\nu)\Psi=0\;,
 \end{eqnarray}
 where $q$ denotes the charge of the scalar field.

 By taking the ansatz of the scalar field
 \begin{eqnarray}
 \Psi=\sum_{l,m}\frac{1}{\sqrt{r(r-2M)}}\phi_{lm}(u, r) Y_{lm}(\theta, \phi)\;,
 \end{eqnarray}
 the Klein-Gordon equation is shown to be separable in this background. After some algebra, we can finally get the following partial differential equation
 \begin{eqnarray} \label{KG}
 &&\partial_v\partial_r \phi+\frac{r-2M}{2r}\partial^2_r \phi
 +\frac{M}{r^2}\partial_r \phi
 +\frac{iqQ}{r}\partial_r \phi
 -\frac{iqQ}{2r^2}\phi\nonumber\\&&
 -\frac{1}{2r^3(r-Q^2/M)}\left(Q^2-2Mr +
 \frac{Q^4}{4 M^2} \frac{r - 2 M}{r - Q^2/M}\right)\phi\nonumber\\&&
 -\frac{l(l+1)}{2r(r-Q^2/M)}\phi=0\;,
 \end{eqnarray}
 where we have dropped the mode number subscripts to simplify the notation.

 We can now evolve the scalar field according to this partial differential equation numerically with the initial condition. From the frequency domain analysis, we know that the mirror-like boundary condition should be imposed to generate the superradiant instability of charged scalar filed in the stringy charged black hole spacetime. This condition now can be explicitly expressed as
 \begin{eqnarray}\label{BC}
 \phi(v, r_m)=0\;,
 \end{eqnarray}
 where $r_m$ is the location of mirror. This condition should be satisfied during the whole time evolution process.

 \section{Numerics}

 We use the method of lines in the forth order Runge-Kutta scheme to integrate in time. The space direction is then approximated by using the pseudo-spectral method instead of the finite difference method. To be more specific, we first use a linear transformation
 \begin{eqnarray}
 z=\frac{2(r-2M)}{r_m-2M}-1\;,
 \end{eqnarray}
 to map the $r$ coordinate interval $[2M, r_m]$ into an interval $[-1, 1]$. We then expand the scalar field in a basis of Chebyshev polynomials in the space direction. The scalar field function is discretized by the values of the expansion on the Gauss-Lobatto collocation points. The spatial derivative of the scalar field is computed by multiplying the Gauss-Lobatto derivative matrix. In the time direction, we use the standard fourth order Runge-Kutta scheme to evolve the scalar field. To be more precise, we first evolve the space derivative of the scalar field according to (\ref{KG}) and then integrate out the scalar field by multiplying the integral matrix with the built in mirror-like boundary condition (\ref{BC}). Therefore the mirror-like boundary condition is automatically maintained in the whole evolution process. The above procedure can be realized in the software of MATHEMATICA. It is noteworthy that the similar numerical trick has also been used in \cite{DTZ} to study the holographic vortex dynamics in the context of AdS/CFT correspondence.

 \section{result}

\begin{figure*}
\subfigure{\includegraphics[width=1.5in]{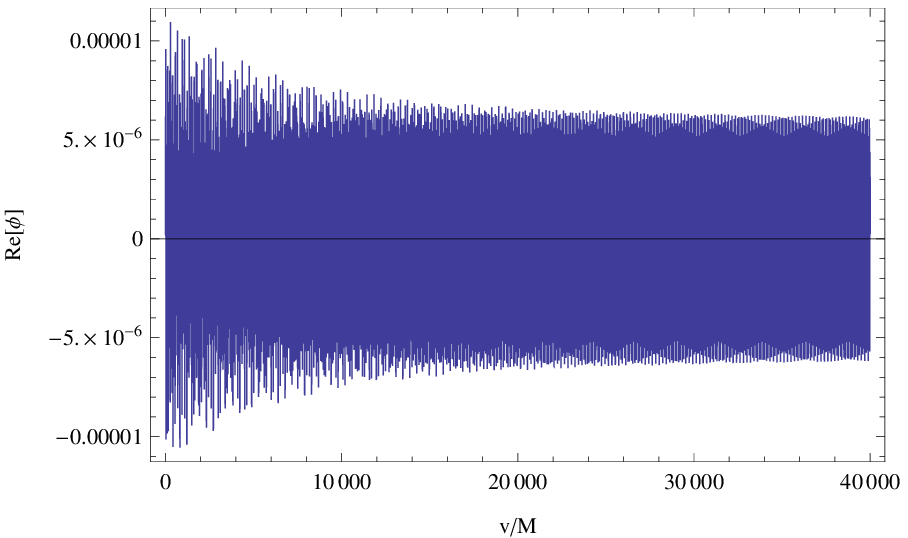}}
\subfigure{\includegraphics[width=1.5in]{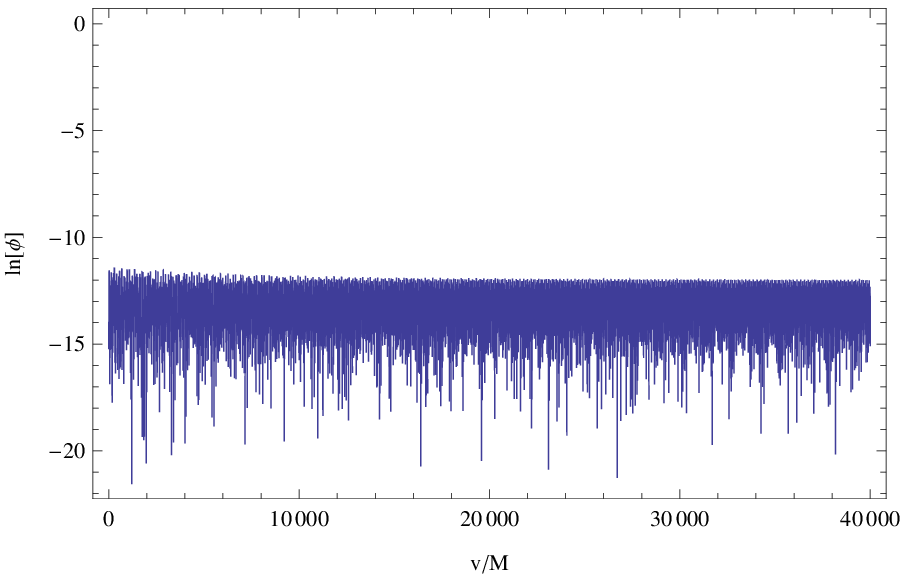}}
\subfigure{\includegraphics[width=1.5in]{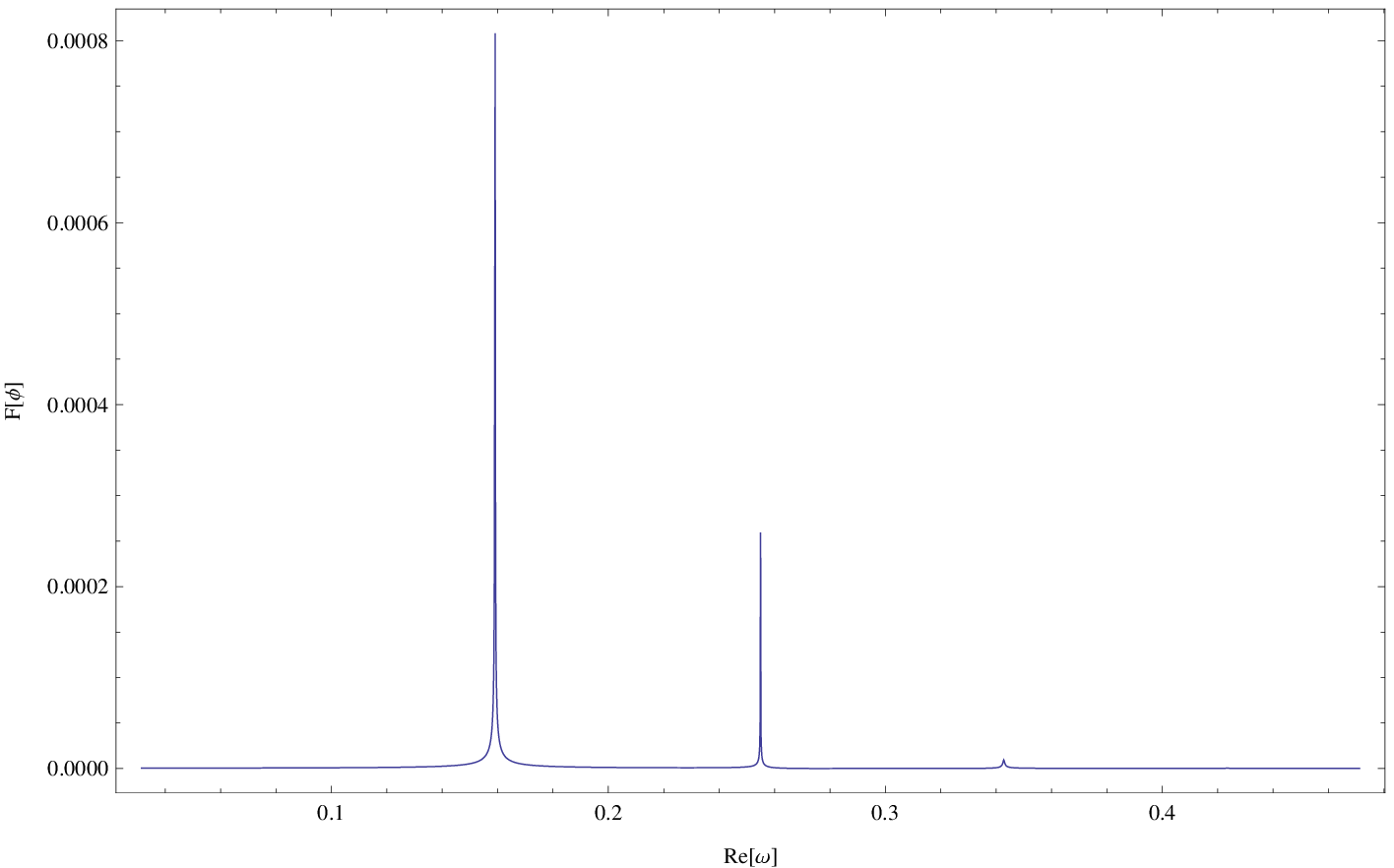}}
\caption{Time evolution of scalar field with the initial Gaussian wave packet. The parameters
are taken as $r_m=30, l=1, Q=0.6$,and $q=0.6$,  with $M=1$ as our unit. The first panel shows the real part of the scalar field evolving with the time, while the second panel shows the logarithm of scalar field evolving with the time. Fourier transformation of the time evolution data is demonstrated in the third panel.}
\end{figure*}

 \begin{table}
 \caption{Boxed quasinormal modes for the first 2 mirrored states
 with $M=1, r_m=30, l=1, Q=0.6, q=0.6$. The second column is taken
 from the frequency domain analysis, while the third column is read off
 from the Fourier transformation of evolution data, i.e. the third panel in Figure 1.}
\centering
\begin{tabular}{lccc}
\hline
\hline
$n$ & $\omega$ & $Re[\omega]$ \\ \hline
$0$\;\; &  $0.158905+5.69687\times10^{-7}i$\;\;   &$0.159$ \\
$1$\;\; &  $0.254763-1.79498\times 10^{-5}i$\;\;  &$0.255$ \\
 \hline
 \hline
\end{tabular}
\end{table}

 Figure 1 shows a generic evolution of scalar field. The initial configuration at $v=0$ is a Gaussian wave packet centered at $r_{cg}$, i.e.,
 \begin{eqnarray}
 \phi(0, r)=\phi_0 e^{-\frac{(r-r_{cg})^2}{\sigma^2}}\;,
 \end{eqnarray}
 with $\phi_0=3\times 10^{-5}$, $r_{cg}=15M$, $\sigma^2=30M^2$. From the first panel in Figure 1, we find that at early stage of evolution the amplitude of scalar field oscillatorily decays with the time. This is mainly caused by the absorption of the black hole horizon. While, in the following stage, it oscillatorily grows very slowly. The reason is that the imaginary part of unstable mode in this case is very small, which is shown in Table I. The second panel shows the logarithm of scalar field amplitude evolving with the time. The third panel is the corresponding Fourier transformation of the evolution data, where some peaks show up at a set of distinct frequencies with the unstable mode dominant. As further demonstrated in Table I, the magnitude of these distinct frequencies is in good agreement with the boxed quasinormal modes in the frequency domain analysis.
 We find the consistency between the modes from two sides. The growth time scale is set by the inverse of the imaginary part. So only after a long time evolution of order of $v/M \sim 10^{7}$ can a significant growth of scalar field amplitude be observed in this case. In what follows we will focus on the case in which the unstable mode has a much bigger imaginary part.

\begin{figure*}
\subfigure{\includegraphics[width=1.5in]{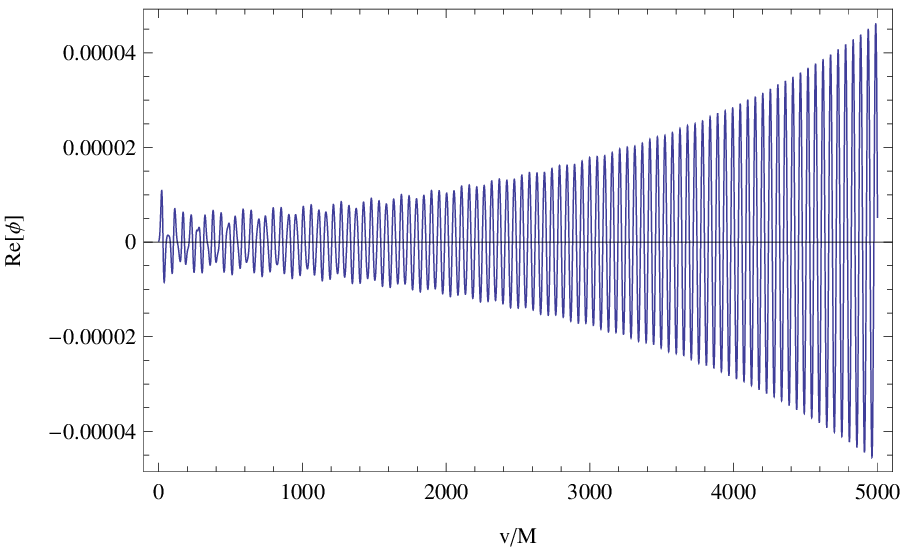}}
\subfigure{\includegraphics[width=1.5in]{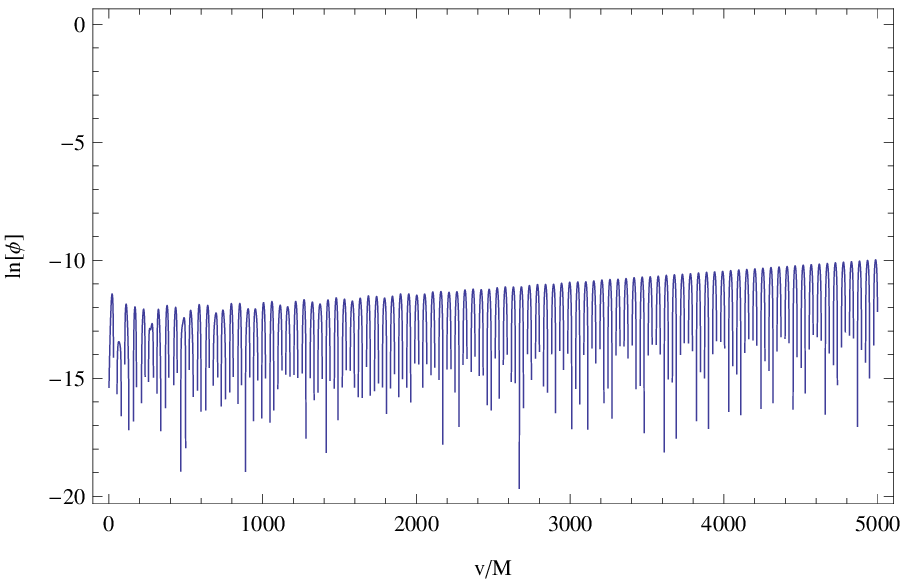}}
\subfigure{\includegraphics[width=1.5in]{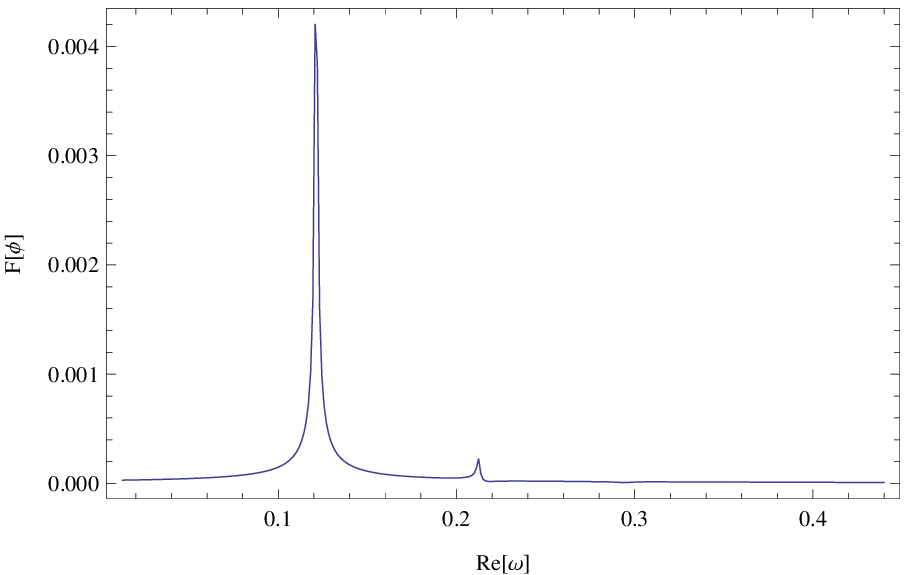}}
\caption{Time evolution of scalar field with the initial Gaussian wave packet. The parameters
are taken as $M=1, r_m=30, l=0, Q=0.6$, and $q=0.6$. The first panel shows the real part of the scalar field evolving with the time, while the second panel shows the logarithm of scalar field evolving with the time. Fourier transformation of the time evolution data is demonstrated in the third panel.}
\end{figure*}

 \begin{table}
 \caption{Boxed quasinormal modes for the first 2 mirrored states
 taking $M=1, r_m=30, l=0,Q=0.6, q=0.6$. The second column is taken
 from the frequency domain analysis, while the third column is read off
 from the Fourier transformation of evolution data, , i.e. the third panel in Figure 2.}
\centering
\begin{tabular}{lccc}
\hline
\hline
$n$ & $\omega$ & $Re[\omega]$ \\ \hline
$0$\;\; &  $0.119962+4.85725\times 10^{-4}i$\;\;   &$0.120$ \\
$1$\;\; &  $0.211098-7.12823\times 10^{-4}i$\;\;  & $0.212$ \\
 \hline
 \hline
\end{tabular}
\end{table}

 By calculating the boxed quasinormal modes, we find that the imaginary parts of $l=0$ unstable modes are generically three orders of magnitude greater than that of $l=1$ unstable modes. This indicates that the $l=0$ unstable modes have a much faster growth rate than the $l=1$ unstable modes, which is confirmed by Figure 2 for the real time evolution of the $l=0$ case. In particular, the exponential growth rate can be estimated as $5\times 10^{-4}$ or so by the time evolution of logarithm of scalar field amplitude, which coincides with the imaginary part of the unstable mode in Table II. Furthermore, as shown in Table II, the real part of unstable mode read off from the Fourier transformation of evolution data is also consistent with the frequency domain analysis.


 In the above two cases, we have chosen the parameters to make the black hole-mirror system possess only one unstable mode. Note that the superradiant critical frequency is given by \cite{dilatonsr}
 \begin{eqnarray}
 \omega_c=\frac{qQ}{2M}\;.
 \end{eqnarray}
 The critical frequency for the above two cases, separating growing modes $(\omega<\omega_c)$ from decreasing modes $(\omega>\omega_c)$, is given by $0.18$, consistent with the above numerical results.
 In order to see more unstable modes show up, it is reasonable to increase the critical frequency by cranking up either the charge of the black hole or scalar field.

 We first compare the time evolutions of scalar field for different values of black hole charge $Q$. As discussed above, in order to observe the distinct growth of scalar field amplitude, we have also chosen the parameter $l=0$. Keeping $M=1,r_m=30$, and $q=0.6$, in the following we take three cases with $Q=0.8, 1.1$, and $1.4$ on top of the previous case of $Q=0.6$. The superradiant critical frequencies in these three cases are $0.24$, $0.33$, and $0.42$, individually. In Figure 3, we display the time evolutions of the scalar field amplitude (left three panels). The first 3 boxed quasinormal modes are displayed in Table III.  As expected before, the number of unstable modes increases with the black hole charge. When $Q=0.8$ and $Q=1.1$, there are two unstable modes. When $Q=1.4$, there are three unstable modes. These unstable modes are all displayed in the Fourier transformation of the evolving data (right three panels). It is also shown that the mode dominates the evolving is always the one with the largest positive imaginary part in Table III. However, there is no general rule of the instability of black hole-mirror system depending on the black hole charge $Q$. In other words, the fast or the slow growth of scalar field is not determined by the large or the small black hole charge.

 \begin{figure}
  \centering
  \subfigure{\includegraphics[width=1.5in]{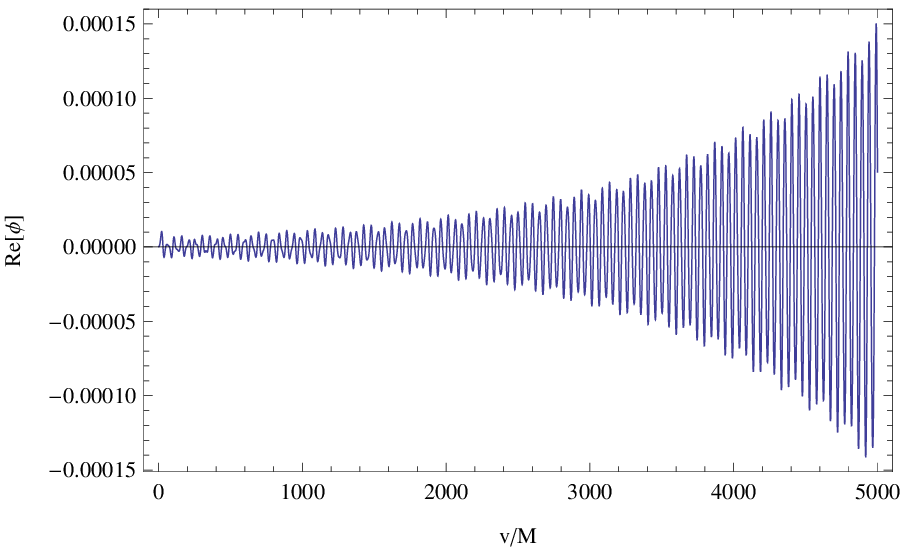}}
  \subfigure{\includegraphics[width=1.5in]{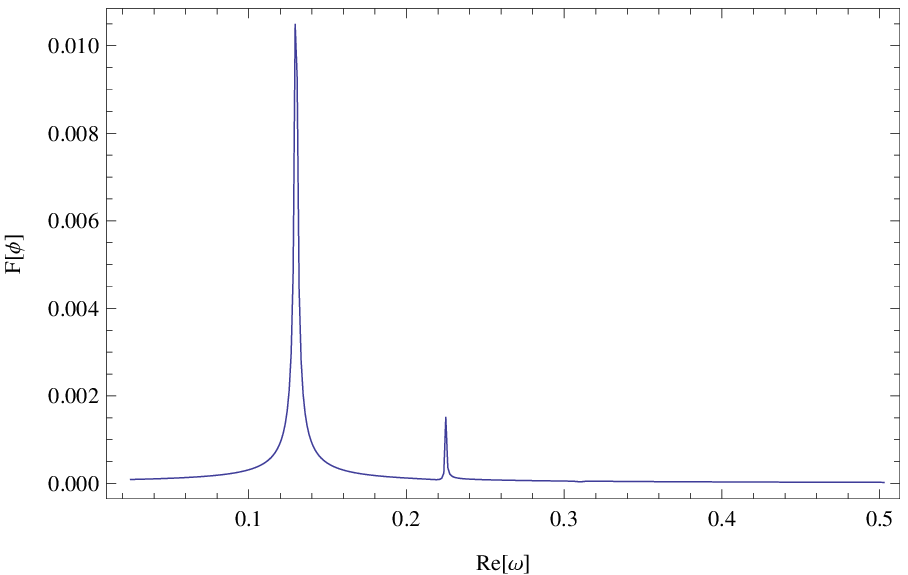}}
 \subfigure{\includegraphics[width=1.5in]{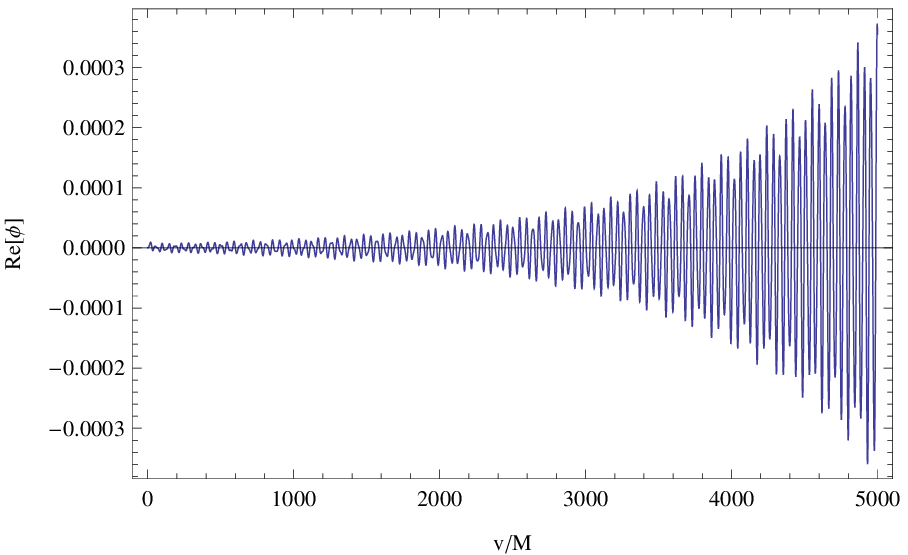}}
 \subfigure{\includegraphics[width=1.5in]{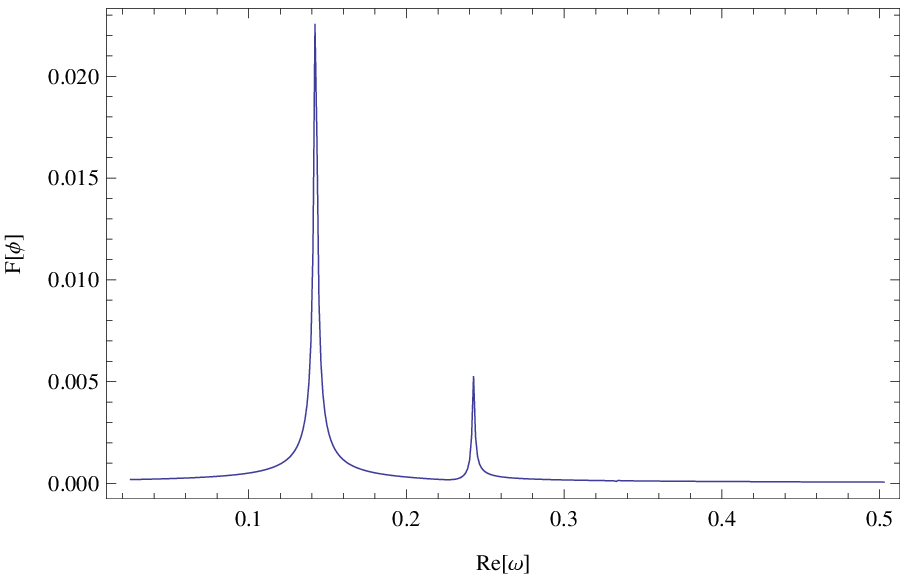}}
  \subfigure{\includegraphics[width=1.5in]{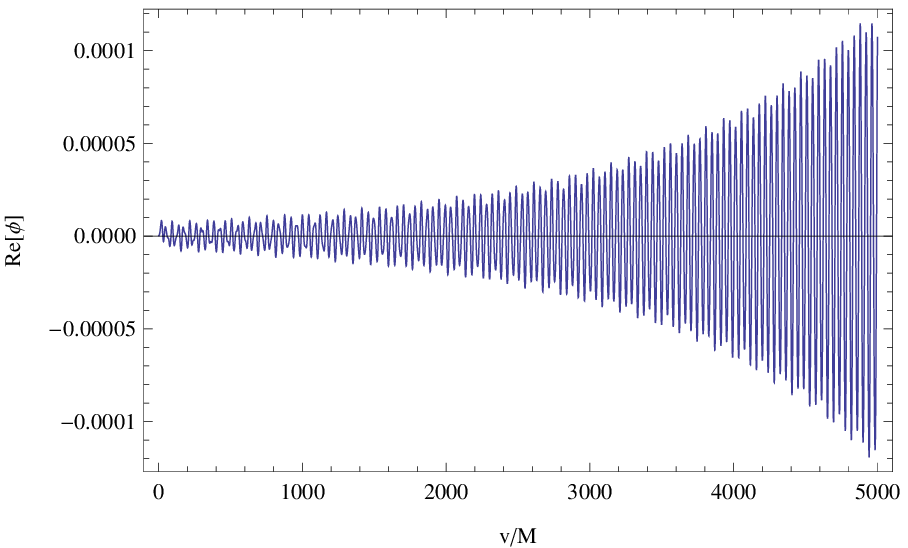}}
    \subfigure{\includegraphics[width=1.5in]{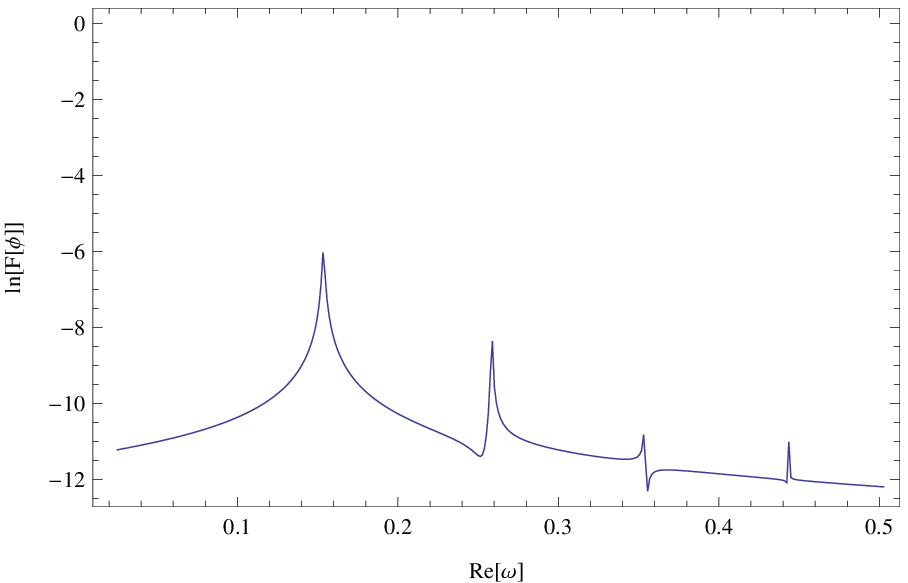}}
   \caption{Time evolutions of scalar field for different black hole charges. From top to bottom, $Q=0.8, 1.1$, and $1.4$, individually. The parameters are taken as $M=1, r_m=30, l=0$, and $q=0.6$. The left three panels show the real part of the scalar field evolving with the time, while the right three panels show the Fourier transformation of the time evolution data. In fact, the last panel shows
   the logarithm of Fourier transformation, otherwise the other modes would be much overshadowed by the most unstable mode.}
\end{figure}

\begin{table*}
 \caption{Boxed quasinormal modes for the first 3 mirrored states
 taking $M=1, l=0, r_m=30, q=0.6$.}
\centering
\begin{tabular}{lccc}
\hline
\hline
$n$ & $Q=0.8$ & $Q=1.1$ & $Q=1.4$ \\
\hline
$0$\;\; &  $0.128718+7.03969\times 10^{-4}i$\;\;   & $0.141096+8.61972\times 10^{-4}i$ \;\;   & $0.15229+6.23948\times 10^{-4}i$ \\
$1$\;\; &  $0.223744+2.33795\times 10^{-4}i$\;\;  & $0.24119+5.9991\times 10^{-4}i$  \;\;   & $0.257199+1.24213\times 10^{-4}i$ \\
$2$\;\; &  $0.309293-2.52535\times 10^{-3}i$\;\;  & $0.331962-2.75849\times 10^{-5}i$  \;\;
  & $0.352634+4.42053\times 10^{-5}i$ \\
 \hline
 \hline
\end{tabular}
\end{table*}

\begin{figure}
  \centering
  \subfigure{\includegraphics[width=1.5in]{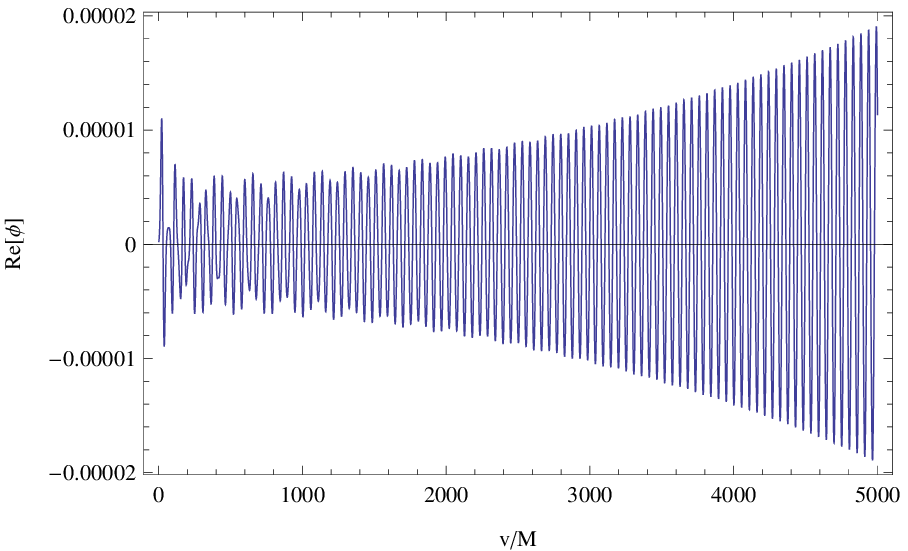}}
  \subfigure{\includegraphics[width=1.5in]{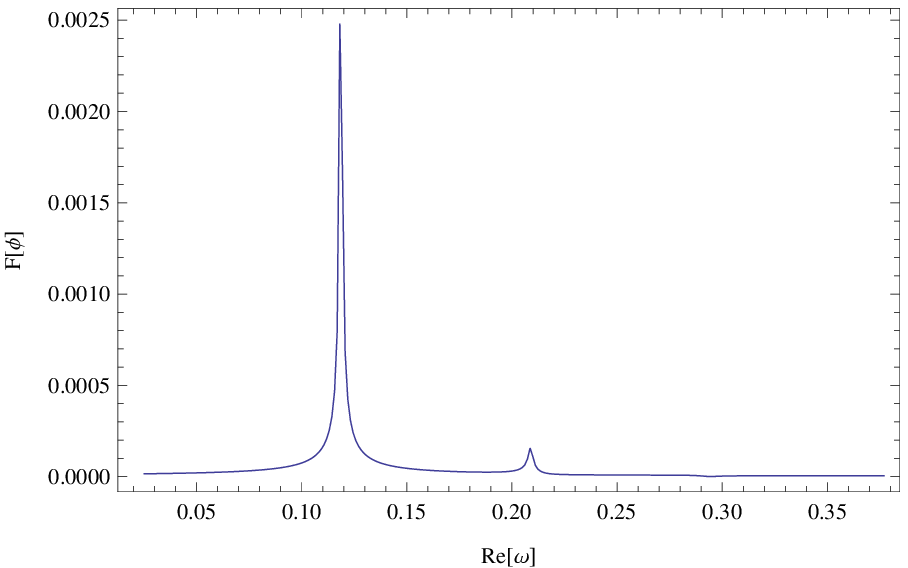}}
 \subfigure{\includegraphics[width=1.5in]{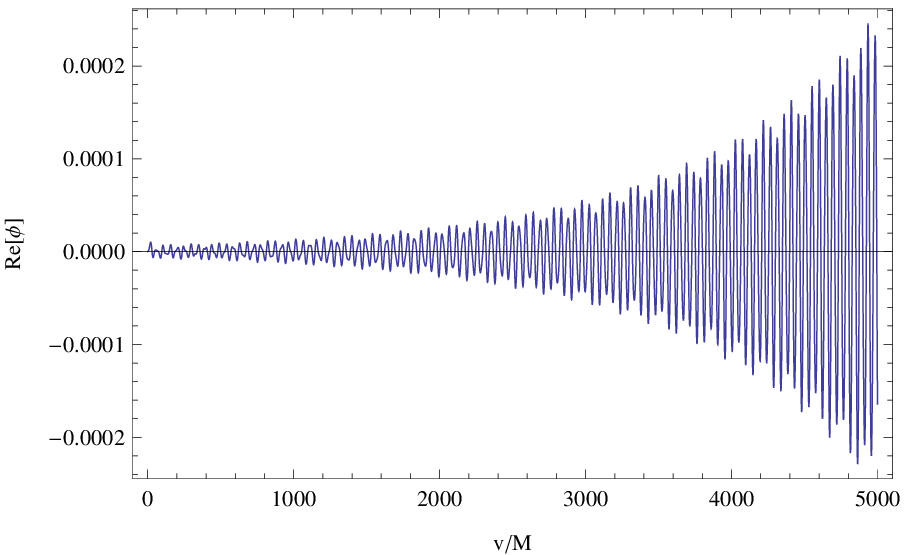}}
 \subfigure{\includegraphics[width=1.5in]{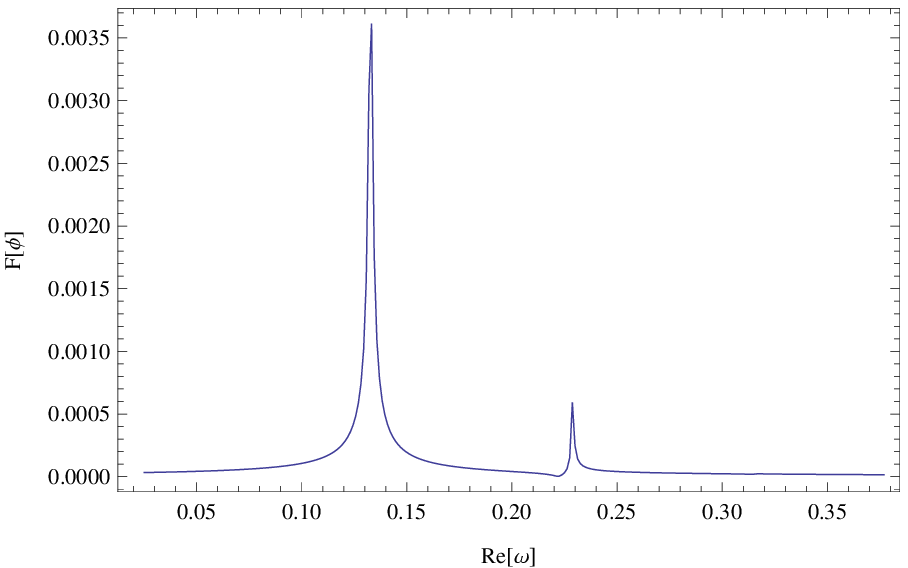}}
  \subfigure{\includegraphics[width=1.5in]{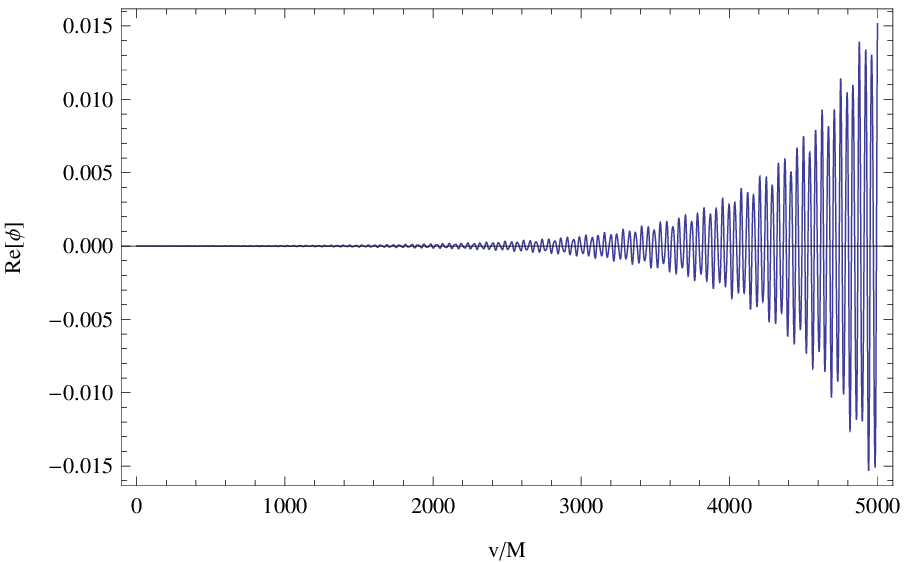}}
    \subfigure{\includegraphics[width=1.5in]{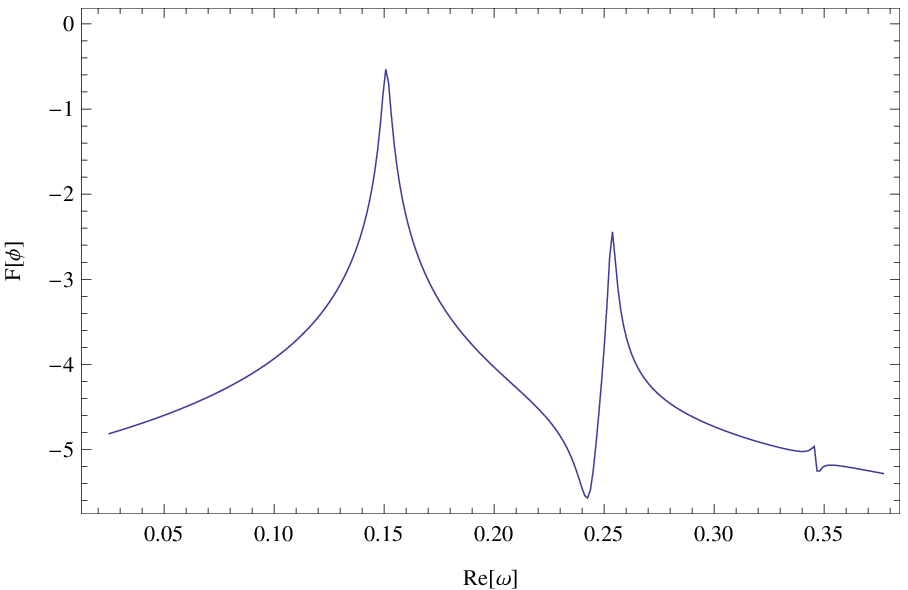}}
   \caption{Time evolutions of scalar field for different values of the field charges. From top to bottom, $q=0.4, 0.65$, $1.0$, individually. The parameters are taken as $M=1, r_m=30, l=0$, and $Q=0.8$. The left three panels show the real part of the scalar field evolving with the time, while the right three panels show the Fourier transformation of the time evolution data. The last panel also shows
   the logarithm of Fourier transformation.}
\end{figure}

\begin{table*}
 \caption{Boxed quasinormal modes for the first 3 mirrored states
 taking $M=1, r_m=30, l=0, Q=0.8$.}
\centering
\begin{tabular}{lccc}
\hline
\hline
$n$ & $q=0.4$ & $q=0.65$ & $q=1.0$ \\
\hline
$0$\;\; &  $0.117342+3.09664\times 10^{-4}i$\;\;   & $0.131452+8.03298\times 10^{-4} i$ \;\;   & $0.149771+1.62917\times 10^{-3}i$ \\
$1$\;\; &  $0.207633-1.04707\times 10^{-3}i$\;\;  & $0.227552+4.26772\times 10^{-4} i$  \;\;   & $0.252281+1.29866\times 10^{-3}i$ \\
$2$\;\; &  $0.289823-7.09925\times 10^{-3}i$\;\;  & $0.314069-1.76193\times 10^{-3} i$  \;\;
  & $0.345103+8.61385\times 10^{-4}i$ \\
 \hline
 \hline
\end{tabular}
\end{table*}

 Next, we compare the time evolutions of scalar field for different values of the field charge $q$, keeping the same initial Gaussian wave packet as before with $M=1, r_m=30, l=0$, and $Q=0.8$. In Figure 4, we display the time evolutions of the field amplitude (left three panels) and the Fourier transformation of the evolution data (right three panels). The boxed quasinormal modes for the first $3$ mirrored states are displayed in Table IV. The superradiant critical frequencies in these three cases are $0.16$, $0.26$, and $0.40$, individually. As expected, the number of unstable modes also increases with the charge of scalar field. When $q=0.4$, there is only one unstable mode. When $q=0.65$, there are two unstable modes. When $q=1.0$, there are three unstable modes. Similarly, the Fourier transformation of the evolution data shows that the dominant mode is always the one with the largest positive imaginary part in Table IV. In addition, consistent with the previous analytic result \cite{liepjc2014}, we also observe that when the field charge $q$ increases, the black hole-mirror system becomes more unstable and exhibits a more rapidly growing behavior.

\begin{figure}
  \centering
  \subfigure{\includegraphics[width=1.5in]{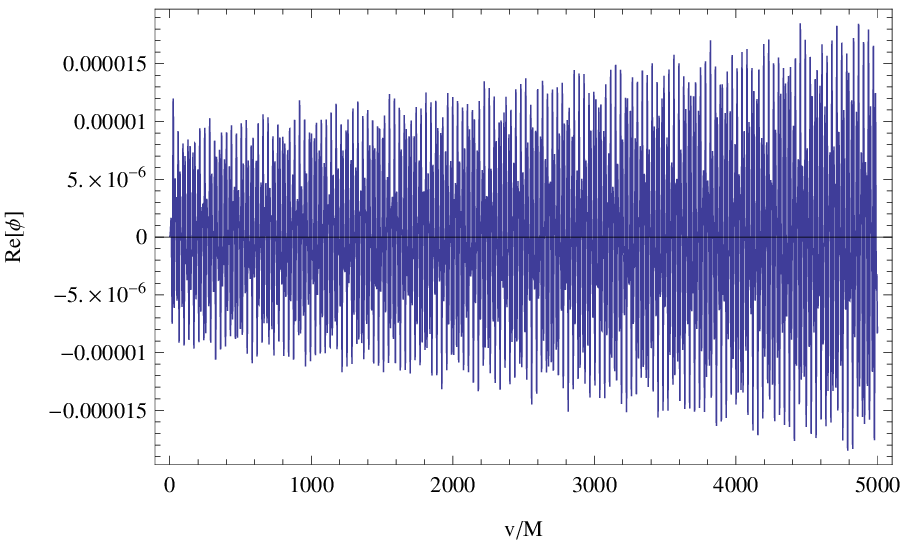}}
  \subfigure{\includegraphics[width=1.5in]{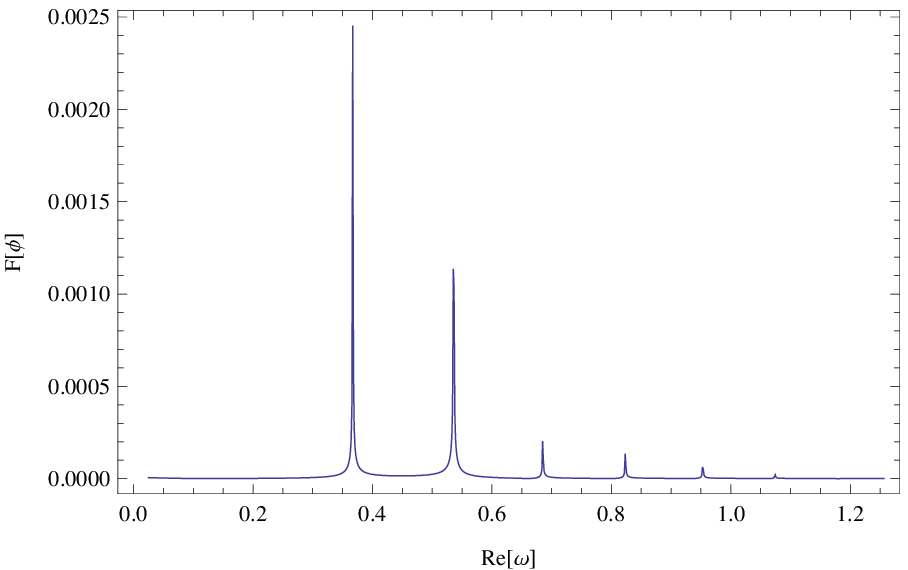}}
 \subfigure{\includegraphics[width=1.5in]{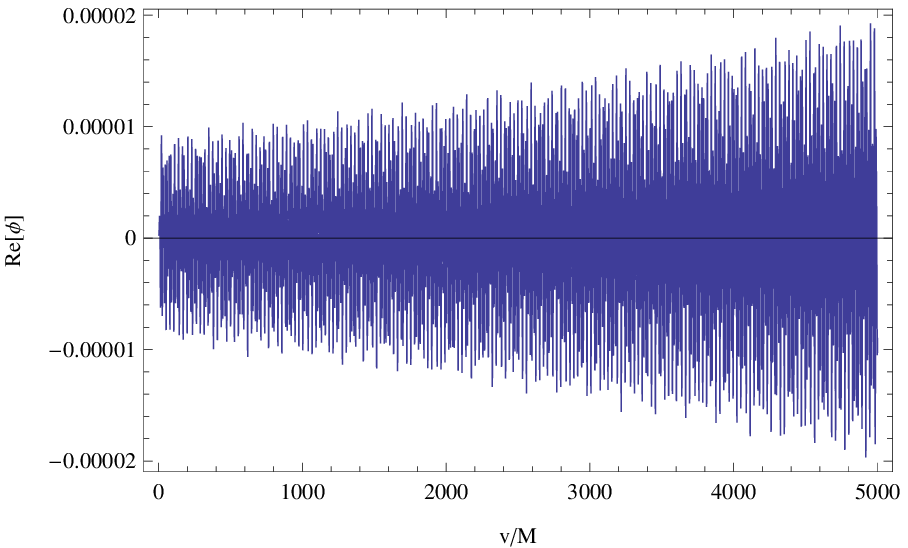}}
 \subfigure{\includegraphics[width=1.5in]{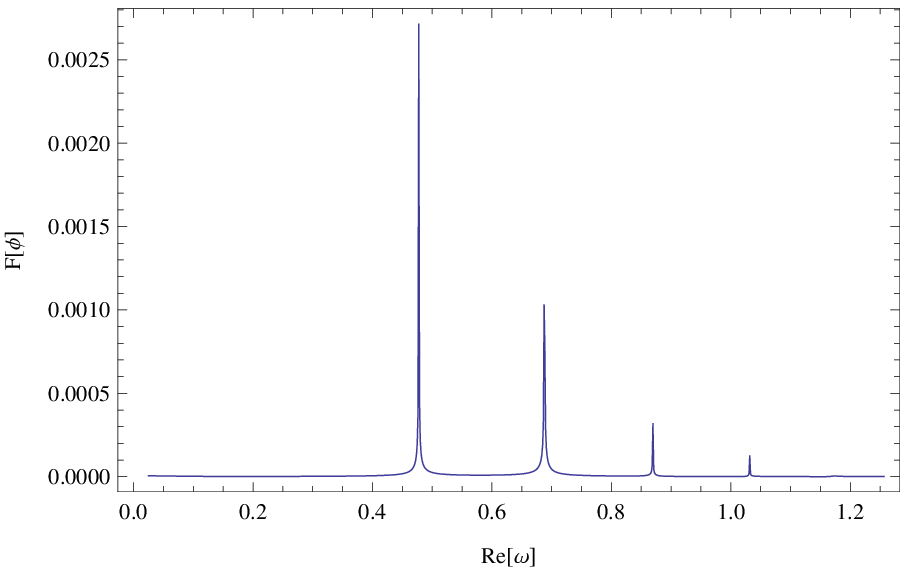}}
  \subfigure{\includegraphics[width=1.5in]{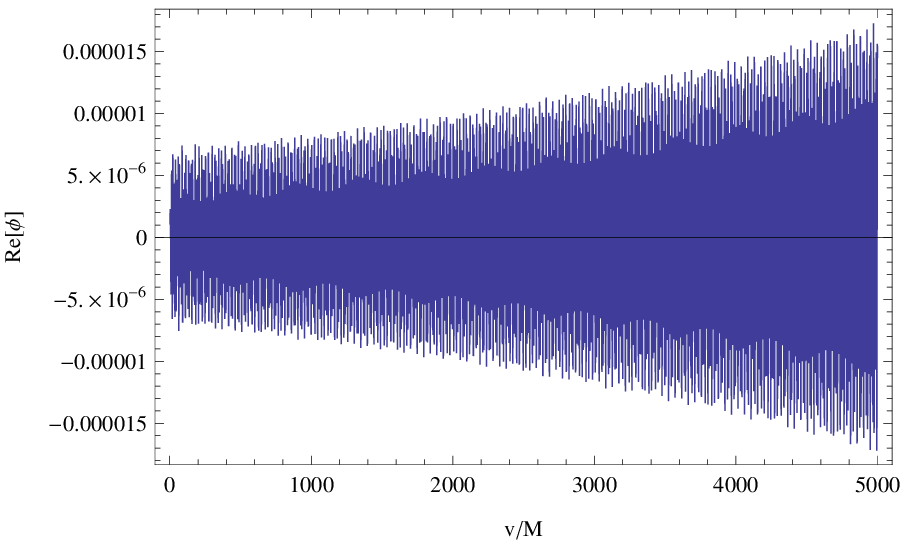}}
    \subfigure{\includegraphics[width=1.5in]{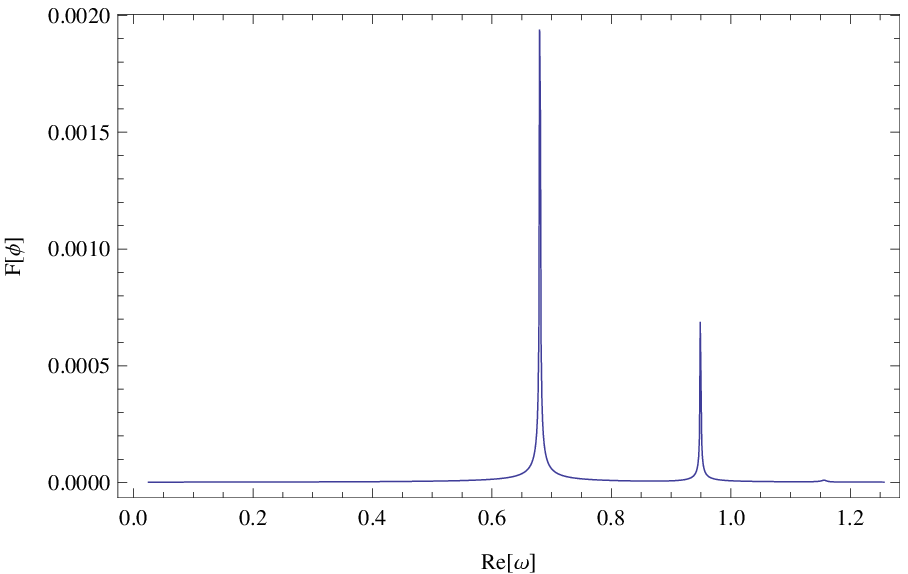}}
   \caption{Time evolutions for different values of mirror radius. From top to bottom, $r_m=20, 15$, and $10$, individually. The parameters are taken as $M=1, l=1, Q=1$, and $q=2$. The left three panels show the real part of the evolving scalar field, while the right three panels show the Fourier transformation of the evolution data.}
\end{figure}

\begin{table*}
 \caption{Boxed quasinormal modes for the first 3 mirrored states
 taking $M=1, l=1, Q=1, q=2$.}
\centering
\begin{tabular}{lccc}
\hline
\hline
$n$ & $r_m=20$ & $r_m=15$ & $r_m=10$ \\
\hline
$0$\;\; &  $0.365882+1.48242\times 10^{-4}i$\;\;   & $0.47635+1.75276\times 10^{-4}i$ \;\;   & $0.67915+2.08701\times 10^{-4}i$ \\
$1$\;\; &  $0.53467+1.13535\times 10^{-4}i$\;\;  & $0.686655+1.29391\times 10^{-4}i$  \;\;   & $0.947995+1.0283\times 10^{-4}i$ \\
$2$\;\; &  $0.684008+9.61229\times 10^{-5}i$\;\;  & $0.868147+1.01999\times 10^{-4}i$  \;\;
  & $1.15524-4.12335\times 10^{-3}i$ \\
 \hline
 \hline
\end{tabular}
\end{table*}

We also show the time evolutions of scalar field for different values of mirror radius $r_m$ in Figure 5. The corresponding boxed quasinormal modes for the first $3$ mirrored states are also displayed in Table V. As expected, Fourier transformation of time evolution data is in good agreement with the frequency domain analysis.

\begin{figure}\centering
  \subfigure[\;]{\includegraphics[width=1.5in]{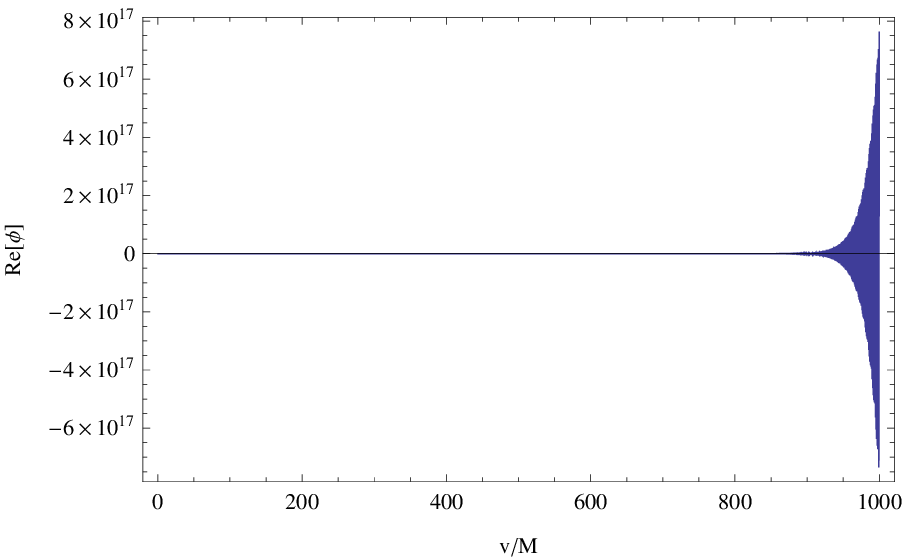}}
  \subfigure[\;]{\includegraphics[width=1.5in]{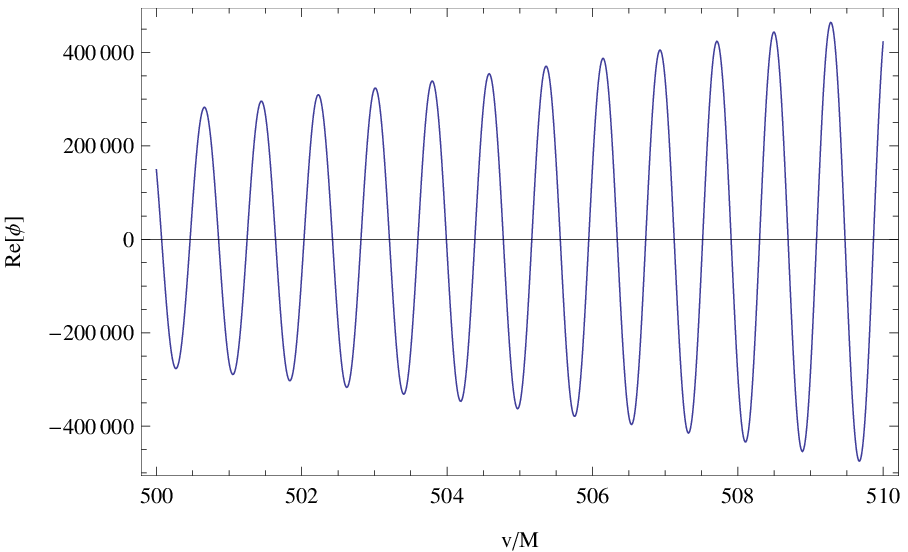}}
 \subfigure[\;]{\includegraphics[width=1.5in]{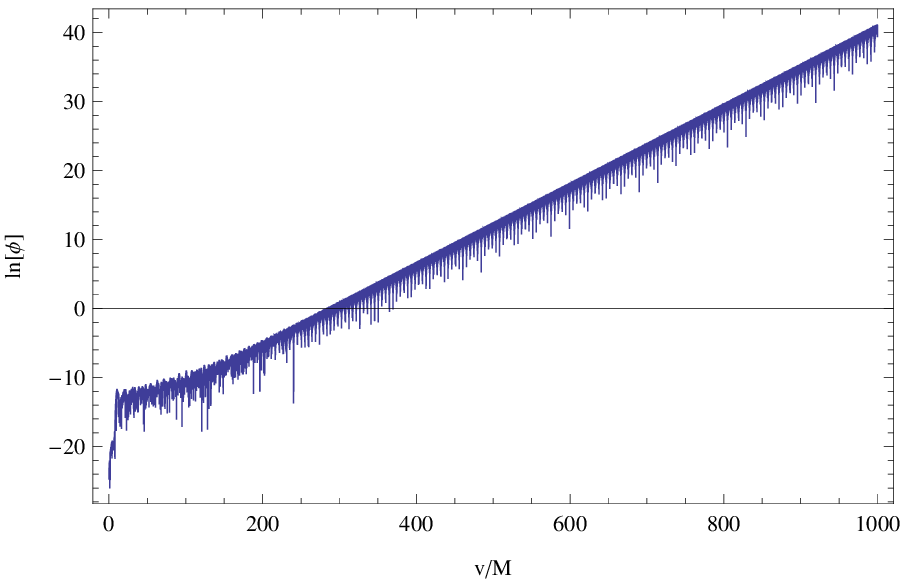}}
 \subfigure[\;]{\includegraphics[width=1.5in]{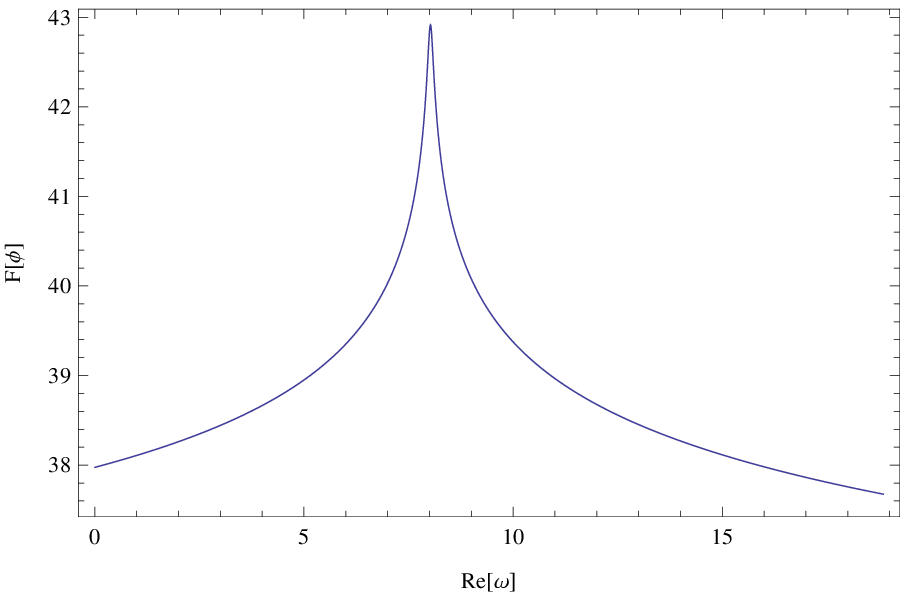}}
\caption{Rapidly growing time evolution. The parameters
are taken as $M=1, r_m=5, l=1, Q=1$, and $q=30$. (a) Real part of the scalar field evolving with the time; (b) Oscillating behavior of real part; (c) Logarithm of scalar field evolving with the time; (d) Logarithm of Fourier transformation of the evolution data.}
\end{figure}

It is shown in the work \cite{degolladoprd2013, degollado} that there exists rapid growth unstable modes in the charged Reissner-Nordstrom black hole. Motivated by this, we would like to end this section by demonstrating a very rapidly growing time evolution in our model. As such,
we take the parameters as $M=1, r_m=5, l=1, Q=1$, and $q=30$, for which the most unstable mode is given by $\omega=8.01818+0.0573857i$. Figure 6 displays such a time evolution of scalar field the Fourier transformation of the evolution data for the initial Gaussian wave packet
 \begin{eqnarray}
 \phi(0, r)=3\times 10^{-5} e^{-\frac{(r - 2.5 M)^2}{0.03 M^2}}\;.
 \end{eqnarray}
 Obviously the real part of scalar field evolves with the time very rapidly. Figure 6(a) and (c)
 represent the growth of the real part of scalar field and its logarithm. It can be easily estimated from Figure 6(c) that the growth rate is about $0.05$, which is in good agreement with the imaginary part of the above unstable mode. In addition, we also zoom in on the time evolution to illustrate the oscillating behavior of the scalar field in Fig. 6(b). To read off the precise value of this oscillating frequency,
 we further take the Fourier transformation of the evolution data and the resultant real part of dominant unstable modes is $8.015$, which is also consistent with the frequency domain analysis as well. Our numerical results show that the rapidly growing evolution process is dominated by the most unstable mode and other unstable modes are much suppressed.

 \section{conclusion}

  In this paper, we have presented a time domain analysis of long time evolution of charged scalar field configuration in charged stringy black hole background by imposing the mirror-like boundary condition. The numerical scheme we have used is pseudo-spectral method in the space direction supplemented with Runge-Kutta method in the ingoing Eddington-Finkelstein time direction. As a result, this numerical scheme turns out to be highly efficient and amenable to a long time evolution. In particular, our numerical results confirm the superradiant instability of scalar field in this background when the black hole is enclosed in a cavity. Furthermore, it is shown that Fourier transforming the evolution data coincides with the unstable modes calculated from the frequency domain analysis. This paper may provide an alternative method to study the time evolution question of field perturbations in black hole background.

  In the present work, the backreaction of scalar field on the background geometry is not considered.
  During the superradiant process, the field extracts energy from the black hole and its energy
  grows exponentially. Then the nonlinear interaction between the field and black hole is required to be taken into account in particular for the case in which there is a very rapidly growing mode. We demonstrate there exists the rapid growth mode exist in charged stringy black hole-mirror system. To see what final state this superradiant instability leads to, the fully nonlinear evolution of the coupled system of scalar field and black hole is required, although it is supposed to much involved. We hope to address this challenging issue in the near future.

 \section*{ACKNOWLEDGEMENT}

  R. L. and J. Z. are supported by NSFC with Grant No. 11205048. R. L. is also supported by the Foundation for Young Key Teacher of Henan Normal University. Y. T. is partially supported by NSFC with Grant No.11475179. H. Z. is supported in part by the Belgian Federal Science Policy Office through the Interuniversity Attraction Pole P7/37, by FWO-Vlaanderen through the project G020714N, and by the Vrije Universiteit Brussel through the Strategic Research Program "High-Energy Physics". He is also an individual FWO fellow supported by 12G3515N.

 \end{document}